\documentclass[sn-mathphys,Numbered]{sn-jnl}

\usepackage{graphicx}%
\usepackage{multirow}%
\usepackage{amsmath,amssymb,amsfonts}%
\usepackage{amsthm}%
\usepackage{mathrsfs}%
\usepackage[title]{appendix}%
\usepackage{xcolor}%
\usepackage{textcomp}%
\usepackage{manyfoot}%
\usepackage{booktabs}%
\usepackage{algorithm}%
\usepackage{algorithmicx}%
\usepackage{algpseudocode}%
\usepackage{listings}%
\usepackage{bm}

\begin{document}
	
	\title[Article Title]{Parallel Diffusion Model-based 
		Sparse-view Cone-beam Breast CT}
	
	
	\author[1]{\fnm{Wenjun} \sur{Xia}}\email{xiaw4@rpi.edu}
	
	\author[2]{\fnm{Hsin Wu} \sur{Tseng}}\email{tseng45@radiology.arizona.edu}
	
	\author[1]{\fnm{Chuang} \sur{Niu}}\email{niuc@rpi.edu}
	
	\author[1]{\fnm{Wenxiang} \sur{Cong}}\email{congw@rpi.edu}
	
	\author[3]{\fnm{Xiaohua } \sur{Zhang}}
	
	\author[3]{\fnm{Shaohua} \sur{Liu}}
	
	\author[3]{\fnm{Ruola} \sur{Ning}}
	
	\author[2]{\fnm{Srinivasan}  \sur{Vedantham}}\email{svedantham@arizona.edu}
	
	\author*[1]{\fnm{Ge} \sur{Wang}}\email{wangg6@rpi.edu}
	
	\affil[1]{\orgdiv{Center of Biomedical Imaging Center, 
			Center for Computational Innovations,
			Center for Biotechnology and Interdisciplinary Studies, and
			Department of Biomedical Engineering, School of Engineering}, \orgname{Rensselaer Polytechnic Institute}, \orgaddress{\city{Troy}, \postcode{12180}, \state{NY}, \country{USA}}}
	
	\affil[2]{\orgdiv{Department of Medical Imaging and Biomedical Engineering}, \orgname{University of Arizona}, \orgaddress{\city{Tucson}, \postcode{85721}, \state{AZ}, \country{USA}}}
	
	\affil[3]{
		\orgname{Koning Corporation}, \orgaddress{\city{West Henrietta}, \postcode{14586}, \state{NY}, \country{USA}}}

	
	\abstract{Breast cancer is the most prevalent cancer among women worldwide, and early detection is crucial for reducing its mortality rate and improving quality of life. Dedicated breast computed tomography (CT) scanners offer better image quality than mammography and tomosynthesis in general but at higher radiation dose. To enable breast CT for cancer screening, the challenge is to minimize the radiation dose without compromising image quality, according to the ALARA principle (“as low as reasonably achievable”). Over the past years, deep learning has shown remarkable successes in various tasks, including low-dose CT especially few-view CT. Currently, the diffusion model presents the state of the art for CT reconstruction. To develop the first diffusion model-based breast CT reconstruction method, here we report innovations to address the large memory requirement for breast cone-beam CT reconstruction and high computational cost of the diffusion model.	Specifically, in this study we transform the cutting-edge Denoising Diffusion Probabilistic Model (DDPM) into a parallel framework for sub-volume-based sparse-view breast CT image reconstruction in projection and image domains. This novel approach involves the concurrent training of two distinct DDPM models dedicated to processing projection and image data synergistically in the dual domains.
		Our experimental findings reveal that this method delivers competitive reconstruction performance at half to one-third of the standard radiation doses. 
		This advancement demonstrates an exciting potential of diffusion-type models for volumetric breast reconstruction at high-resolution with much-reduced radiation dose and as such hopefully redefines breast cancer screening and diagnosis.}

	\keywords{Breast CT, image reconstruction, diffusion model, parallel computing, dual-domain processing}
	
	
	
	\maketitle
	
	\section{Introduction}\label{sec1}
	
	Breast cancer remains the most prevalent cancer among women. In 2023, it is estimated that approximately 298,000 women in the United States will be diagnosed with this disease \cite{american2012cancer}. The incidence of breast cancer has been witnessing a gradual increase by about 0.5\% annually since the mid-2000s \cite{siegel2023cancer}. This trend indicates that breast cancer will continue being a primary concern in the coming years.
	
	Encouragingly, the mortality rate for breast cancer is comparatively lower than that for other cancers \cite{american2012cancer, siegel2023cancer, berry2005effect}. This can be largely attributed to the effectiveness of early detection, which plays a crucial role in reducing the mortality rate associated with breast cancer. Among the noninvasive screening methods, mammography, tomosynthesis and breast computed tomography (CT) stand out as the primary modalities for early detection. Notably, mammography is considered the gold standard in breast cancer screening due to its dose-efficiency, effectiveness and accessibility \cite{miller2001breast, poplack2000mammography, linver1997mammography}.
	
	Despite great utilities of mammography, its projective nature remains a significant drawback. This results in overlapped structures, reducing the accuracy of lesion detection. Statistically, about 12\% of women undergoing mammography screening receive abnormal results, necessitating further tests; yet, only approximately 4\% of these cases are confirmed to have cancer \cite{american2012cancer}. While tomosynthesis adds 3D information to some degree \cite{nicosia2023breast}, ideally 3D imaging is always the holy grail as in other medical imaging tasks. For this purpose, breast cone-beam computed tomography (CT) offers 3D images directly. This specialized CT technology enables accurate representation of breast tissues and structures, particularly beneficial for detecting small lesions. Modern breast CT systems boast an image resolution of 100-200 \textmu m, enabling precise lesion identification and localization \cite{yang2007dedicated}. Additionally, breast CT scans are often more comfortable for patients, as they do not require breast compression \cite{lindfors2008dedicated}. With its isotropic resolution, increased accuracy, and patient comfort, breast CT is poised to revolutionize breast cancer imaging \cite{boone2006breast}. As the technology evolves and costs decrease, breast CT is likely to emerge as the new standard in breast cancer screening.
	
	However, the primary challenge with breast CT is the associated radiation exposure. Given the fact that X-ray radiation can induce cancer, there has been significant concern about the risk of CT scans to the patient. The ALARA (''as low as reasonably achievable'') principle is to mitigate this risk \cite{slovis2002alara}. Studies show that the median mean glandular dose (MGD) from breast CT exams is 3 to 4 times higher than that from standard digital mammography \cite{boone2001dedicated, vedantham2013personalized, hendrick2010comparison}, exceeding the limits set by the Mammography Quality Standards Act (MQSA). Addressing this challenge involves reducing the radiation dose of breast CT scans to about 1/2 to 1/3 of the current levels, aligning with acceptable standards for regular screening. This reduction is technically feasible, with sparse-view breast CT being an emerging method of choice \cite{tseng2020sparse}. Sparse-view breast CT involves reducing the number of views captured during a scan to minimize radiation dose. The key is to maintain high-quality imaging with under-sampled projections, which places high demands on image reconstruction algorithms.
	
	Current commercial breast CT scanners primarily use the Feldkamp-Davis-Kress (FDK) algorithm for cone-beam image reconstruction \cite{feldkamp1984practical}. This algorithm requires a circular full scan and tends to produce image artifacts and noise in a sparse-view acquisition geometry, compromising cancer screening and diagnosis. Over the past years, various algorithms have been developed for sparse-view CT reconstruction \cite{lohvithee2017parameter, zeng2013iterative, niu2014sparse, sohn2020analytical}, with deep learning techniques leading the recent advancements \cite{chen2017low, chen2017lowdose, jin2017deep, yang2018low, chen2018learn, adler2018learned, gupta2018cnn, xiang2021fista, chun2020momentum, hu2020hybrid, zhang2021clear}. The popular deep learning methods often employ convolutional neural networks (CNNs) for sparse-view CT reconstruction and need further development to meet clinical needs \cite{wang2016perspective, wang2020deep}.
	
	A typical breast CT scan produces a large projection dataset and then a large image volume, either of which is a 3D dataset challenging to a deep reconstruction workflow on a modern computer workstation. Specifically, each image volume consists of several hundred 2D slices of 1024x1024 pixels. This size makes direct network processing of the entire dataset impractical due to the enormous storage requirements. Currently, the most methods involve processing 2D slices individually \cite{fu2020residual} and do not fully utilize the 3D contextual information. Moreover, direct network post-processing of images can result in the loss of structural details. Therefore, it's highly desirable to directly process projection and image volumetric data in the deep reconstruction workflow \cite{xia2023physics} to handle the cone-beam geometry of breast CT at minimized radiation dose.
	
	In the image reconstruction field, the diffusion models have recently become the state of the art.  In particular, the denoising diffusion probabilistic model (DDPM)-based deep learning models are recognized as a potential solution for image processing, generation, and reconstruction \cite{ho2020denoising, song2020score, xia2022patch, xia2022low}. These models are well established in their commercial applications in Midjourney, Stable Diffusion, and DALL-E 3 for generative AI. The latest integration of DALL-E 3 with ChatGPT, providing a text-guided image generation API, underscores their amazing capabilities. The effectiveness of diffusion models in generating lifelike images suggests their potential applicability in meeting the quality requirements for breast CT image reconstruction.
	
	With its powerful ability to transform normally distributed noise into a desirable image through a series of variational autoencoders (VAEs) \cite{kingma2013auto}, DDPM excels in image quality and training stability compared to generative adversarial networks (GANs) \cite{goodfellow2020generative}. Its superiority over other generative models has been demonstrated in various applications such as image super-resolution \cite{rombach2022high, saharia2022image}, image inpainting \cite{lugmayr2022repaint}, and image editing \cite{meng2021sdedit}. The successes of DDPM were also reported in the CT field \cite{xia2022low, li2022ultra, peng2023cbct}, surpassing traditional networks in terms of reconstruction quality. However, applying DDPM to breast CT reconstruction must address the large size of breast CT data and image data and the intensive computational cost of the diffusion process.
	
	In this study, for the first time we propose to parallelize the DDPM for breast cone-beam CT reconstruction. Our method strategically divides the image reconstruction process into parallel sub-processes in both the projection and image domains. This division is based on handling sub-volumes of projection and image data, which are processed by two distinct DDPMs. Our approach first assembles these sub-volumes into a comprehensive dataset of cone-beam projections. Then, we employ the Feldkamp-Davis-Kress (FDK) algorithm for domain transformation, and extract sub-volumes of the resultant image volume for further refinement via DDPM again.
	
	A main feature of our methodology is the independence of these sub-volumes, allowing us to implement a parallel processing pipeline in a distributed computing system. This structure necessitates only minimal data broadcasting during the domain transformation phase, markedly enhancing the overall efficiency of the process. Our key contributions are twofold, collectively marking a notable advancement in deep learning-based breast CT reconstruction:
	\begin{itemize}
		
		
		\item \textbf{Innovative Method for Volumetric Reconstruction of High-resolution CT Images from Sparse Projections:} We have developed an innovative approach to address the challenges associated with applying the diffusion or diffusion-type models to reconstruct large-size high-resolution breast CT images. This method effectively decomposes overwhelming memory and computational demands into manageable pieces for parallel processing, overcoming the barrier in the application of cutting-edge diffusion models to high-end medical image reconstruction. Although this study is focused on cone-beam breast CT, the same approach can be extended to high-resolution photon-counting CT as well, which is the next generation of medical CT technology.
		
		\item \textbf{Clinical Validation and Potential Impact on Women’s Health:} Crucially, our method and results have undergone validation by radiologists, affirming the clinical value of our high-quality reconstruction of breast CT images. This validation highlights the potential of our approach to enhance regular breast cancer screening at much-reduced radiation dose. The improved accuracy and efficiency of our method have the potential to significantly contribute to the improvement of women's health by facilitating earlier and more precise detection of breast abnormalities with minimized ionizing radiation risk.
		
	\end{itemize}
	
	\section{Results}\label{sec2}
	
	In this study, we applied our proposed sub-volume-based parallel DDPM pipeline and achieved the state-of-the-art (SOTA) reconstruction performance for sparse-view cone-beam breast CT. We conducted a comparative analysis between our parallel dual-domain DDPM (DDPM-D), parallel single-domain DDPMs in either the projection (DDPM-P) or image domain (DDPM-I), and an image-domain U-Net model \cite{jin2017deep}. These models are referred to as DDPM-D, DDPM-P, DDPM-I, and U-Net respectively.
	
	In our approach, DDPM-P exclusively utilizes a single DDPM in the sinogram domain to inpaint projection data, subsequently applies the Feldkamp-Davis-Kress (FDK) algorithm for reconstruction. DDPM-I, conversely, operates in the image domain, processing the FDK reconstructed images obtained from sparse-view data. In the image domain, the U-Net model was trained to learn the mapping from sub-volumes of a degraded reconstructed image volume to the high-quality counterparts.
	
	The comparative study was carried out on the datasets emulating two levels of radiation dose: half dose and one-third dose. The fully sampled projection data comprised 300 views, with the half-dose and one-third dose datasets consisting of 150 and 100 views respectively. We evaluated the methods through visual inspection. The image quality in our study was comprehensively evaluated by displaying slices along three axes. This approach allowed for a detailed assessment of the visual quality of the images.
	
	\begin{figure*}[htbp]
		\centering
		\includegraphics[width=1.0\textwidth]{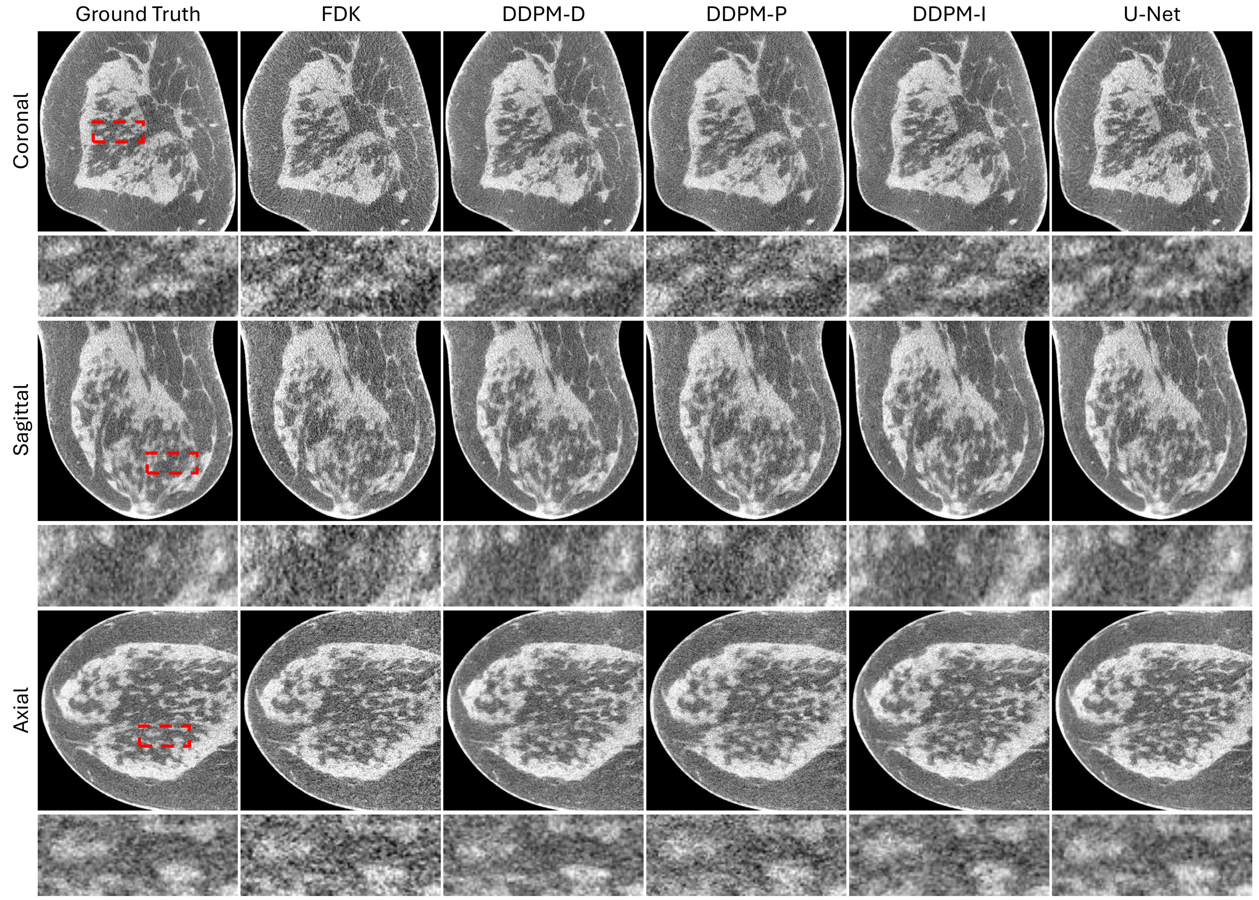}
		\caption{Visualization of the three orthogonal planes through another 3D breast CT volume. The ground truth was reconstructed from fully-sampled projection data (300 views) using the FDK algorithm, and the remaining results were reconstructed from half dose data (150 views) using the competing methods respectively. The display window is set to [-100, 550] HU.}
		\label{fig:4}
	\end{figure*}
	
	\subsection{Half Dose Results}
	
	Fig. \ref{fig:4} presents a typical case of 3D breast CT images across three orthogonal planes respectively. The ground truth was reconstructed from a full-scan dataset (300 views) using the FDK algorithm. In contrast, the other presented results were reconstructed from half-dose data (150 views) using the competing methods.
	
	The few-view FDK reconstruction was contaminated with image noise and artifacts, which adversely affect image quality and thus diagnostic accuracy. It is observed across the three planes that the diffusion models effectively mitigate these artifacts. However, the U-Net results do not completely eliminate them. This is particularly evident in the coronal views where artifacts pervade the entire image, and in the axial views where artifacts are visible near the right edge. Due to its reliance on convolutional neighborhood weighting, it is not surprising that U-Net is more adept at reducing noise than removing artifacts.
	
	Regarding the results of DDPM-P, noise introduced during the processing in the projection domain is noticeable. This noise, though subtle, can be propagated through the FDK algorithm, degrading the overall image quality. Applying DDPM to these initial DDPM-P results effectively removes such noise, as demonstrated in the superior image quality of the DDPM-D results. Both DDPM-D and DDPM-I, utilizing the DDPM model in the image domain, yield results that are remarkably similar in quality and closely resemble the ground truth.

	For an in-depth evaluation, we focused on the area marked with the red dashed box. Here, the DDPM-P results contain pronounced noise, leading to compromised image quality. While U-Net effectively removes noise, this approach may not always align with clinical preference. Radiologists often interpret images based on noise patterns, and excessive noise removal can blur subtle image features and texture, potentially reducing image resolution and diagnostic value. The DDPM-I results, while presenting clearer gland structures, show slight deformation of glands. The DDPM-D results stand out as superior. The dual-domain DDPM not only accurately restores the gland structure but also effectively eliminates noise and artifacts, thereby yielding a superior image reconstruction. 

	

	\subsection{One-third Dose Results}
	
\begin{figure*}[htbp]
		\centering
		\includegraphics[width=1.0\textwidth]{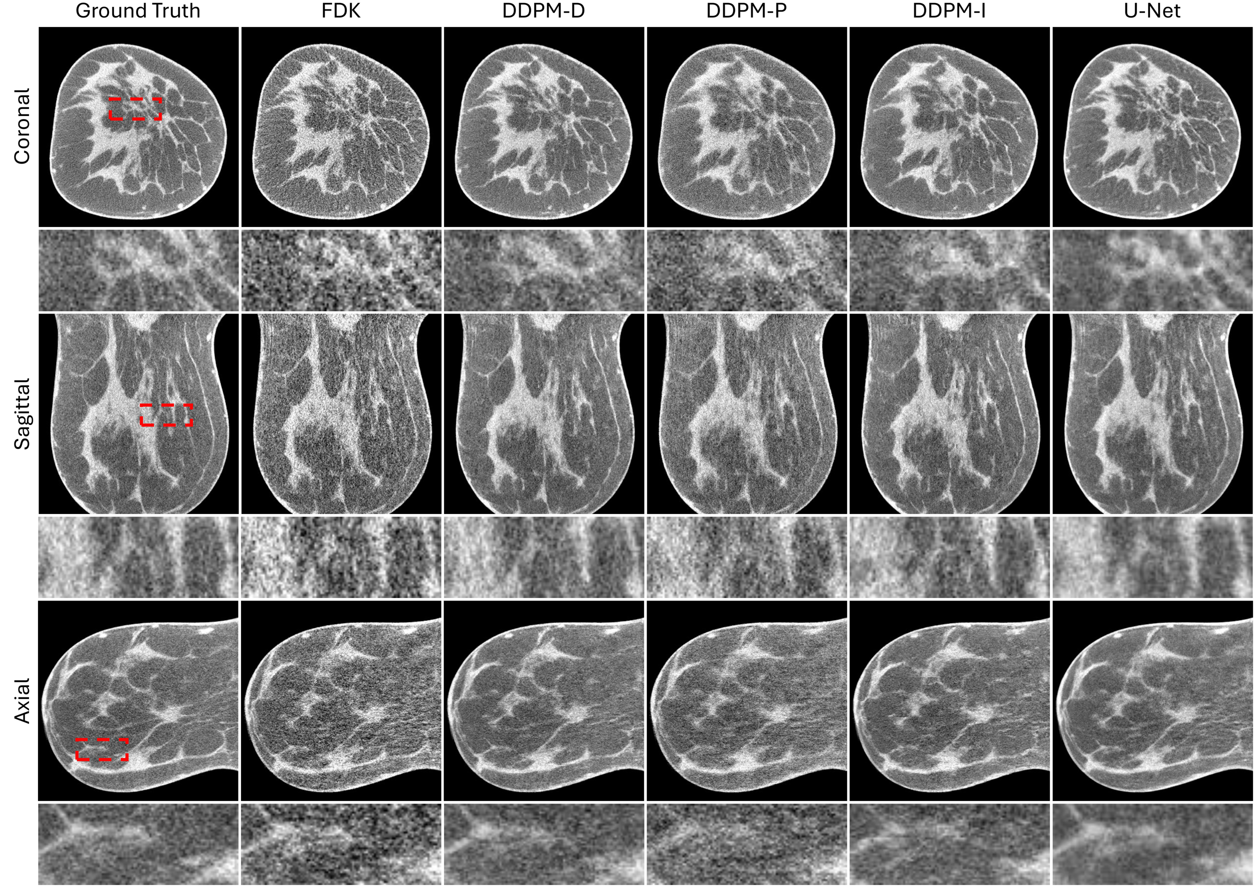}
		\caption{Visualization of the three orthogonal planes through a 3D breast CT volume. The ground truth was reconstructed from fully-sampled projection data (300 views) using the FDK algorithm, and the remaining results were reconstructed from one-third dose data (100 views) using the competing methods respectively. The display window is set to [-100, 550] HU.}
		\label{fig:6}
	\end{figure*}

	Fig. \ref{fig:6} presents the reconstruction results from one-third dose data, juxtaposed with the ground truth reconstructed from fully-sampled data  using the FDK algorithm. Notably, the FDK reconstructions from the one-third dose data are marred by more severe noise and artifacts than those from half-dose data, thus posing a significant challenge to the performance of each algorithm. In the U-Net results, the presence of artifacts, combined with a blurrier image quality, suggests a compromise in image sharpness for agressive noise reduction. In contrast, the DDPM-P results continue being adversely affected by noise, compromising the overall image quality. A comprehensive examination of the images reveals that the diffusion-based methods generally outperform U-Net, particularly in terms of suppressing noise and artifacts. Despite DDPM-P's struggle with noise, it still excels over U-Net that produces strong artifacts and could reduce diagnostic accuracy.
	
	Focusing on the ROIs provides a perspective on how each method to remove noise and artifacts. The U-Net results, although less noisy, show diminished sharpness and exhibit a piecewise smoothing effect on gland structures, diverging from the ground truth. The images from DDPM-P display a higher noise level than their half-dose counterparts, leading to a less clear image presentation. The DDPM-I results, while maintaining clarity, show more distorted gland structures than their half-dose counterparts across all the three orthogonal planes. In this context, the dual-domain DDPM-D method stands out for its exceptional performance in suppressing noise and artifacts while preserving gland structures. Its results not only surpass other methods in terms of restoring gland structure but also closely match the ground truth visually, making it the most preferable choice among the tested methods.
	
	\section{Discussions and Conclusion}\label{sec3}
	
	The breast CT imaging technology holds a great potential to become the new modality for breast cancer screening and diagnosis. However, advancing breast CT from a diagnostic tool to a routine screening alternative necessitates a critical step for minimizing radiation dose and associated patients' risks. This can be achieved by employing sparse-view cone-beam breast CT techniques.
	
	Sparse-view cone-beam breast CT requires advanced image reconstruction algorithms with deep learning, particularly diffusion models as the state-of-the-art generative AI approach. As a solid starting point, the DDPM model has emerged for imaging tasks including CT image reconstruction. However, applying DDPM to breast CT reconstruction meets two major challenges: the breast CT volume is rather large owing to its high image resolution, and the generic diffusion process takes thousands of iterations taking prohibitively long computational time. In this work, we have successfully tackled both the challenges by implementing a parallel DDPM model in the dual-domains. Our experimental results demonstrate the superiority of our method. Remarkably, the parallel dual-domain DDPM consistently outperforms traditional reconstruction methods, making similar advancements observed in other fields \cite{dhariwal2021diffusion}.
	
	Our diffusion-based reconstruction pipeline, designed for independent sub-volumes in parallel, is ideally suited for deployment in a distributed computing system. This approach is consistent with the ongoing evolution of deep learning, where large models increasingly benefit from cloud  platforms. In the future, deep learning applications are expected to be predominantly supported by cloud and edge computing infrastructures, obviating the need for individuals or entities to maintain their own heavy computational resources. Services like ChatGPT are already embracing this trend, and we anticipate a rapid expansion of cloud platforms in the coming years. Specifically, medical equipment reliant on deep learning will likely be integrated with cloud services, taking advantage of their power and flexibility. In this context, the parallelization of algorithms becomes indispensable. The parallel pipeline we propose in this paper serves as an initial model for powerful diffusion model-based tomographic imaging tasks, and can be extended to other cases such as high-resolution photon-counting CT.
	
	While DDPM shows promise, its major drawback is the time-consuming inference process, typically consisting of thousands of diffusion steps. This has been the bottleneck for the DDPM model, making acceleration techniques critical \cite{song2020denoising, lu2022dpm, lu2022dpm++, xia2022low}. Our work is distinguished as the first to parallelize the diffusion model for distributed, rapid inference, with few-view breast CT as a primary application. As the next step, we can accelerate our parallel DDPM model using emerging techniques including the consistency model and the Poisson flow generative model \cite{song2023consistency, xu2022poisson}.
	
	In conclusion, we have presented the first parallel algorithm for diffusion model-based tomographic reconstruction, addressing significant challenges in enabling sparse-view cone-beam CT for breast cancer screening and diagnosis. Our contributions mark a significant advancement of low-dose cone-beam breast CT technology, paving the way for safer, better and more efficient breast cancer imaging and improving woman's health significantly.
	
	\section{Methods}\label{sec4}
	
	\begin{figure*}[htbp]
		\centering
		\includegraphics[width=1.0\textwidth]{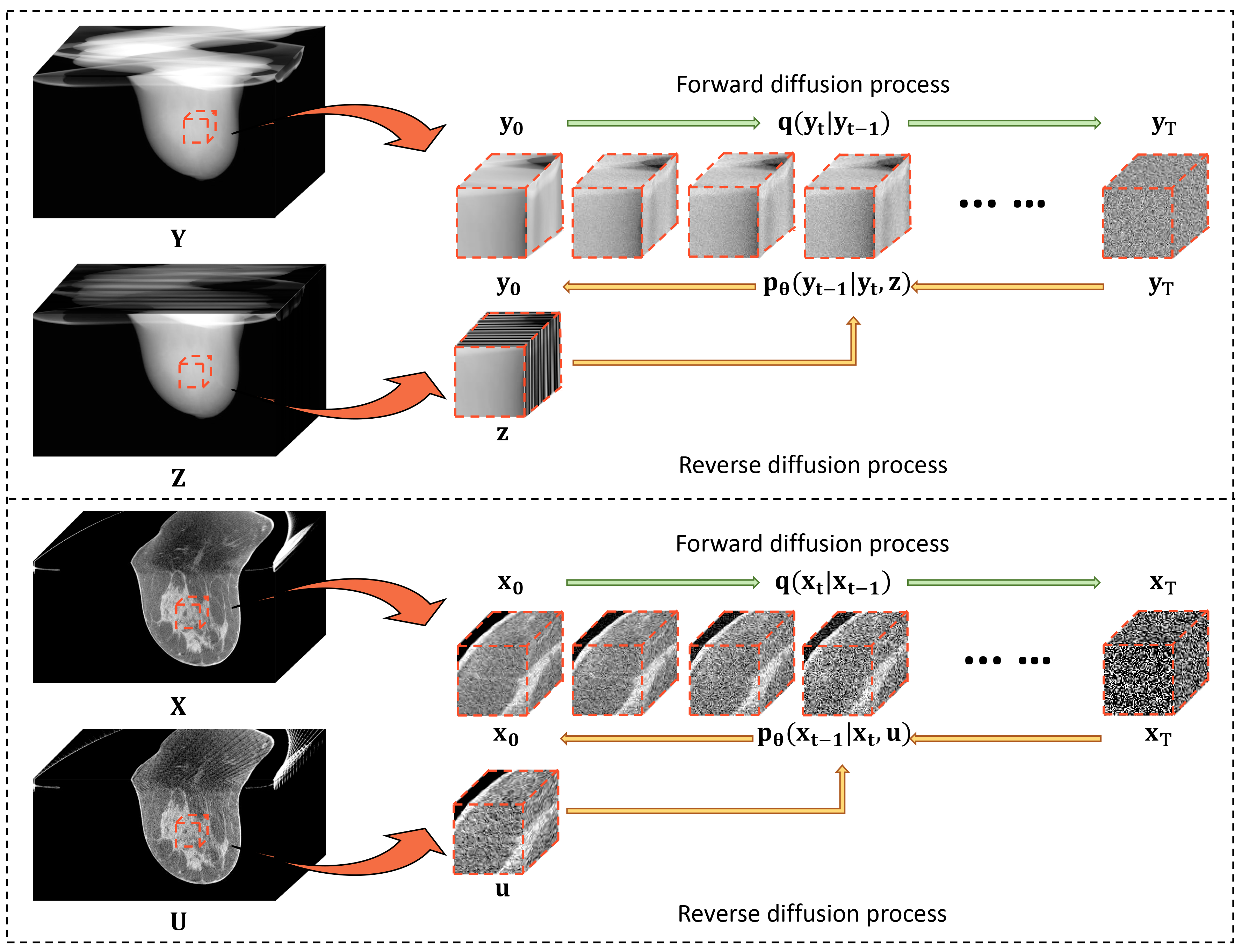}
		\caption{Conditional dual-domain DDPM adapted for sparse-view breast CT.}
		\label{fig:1}
	\end{figure*}
	
	\subsection{Conditional DDPM}
	For the fully sampled breast CT projection data $\bm{Y}$ and the corresponding few-view projection data $\bm{Z}$, the DDPM consists of a forward process and a reverse process. Given a time series $ \{1, 2, ...,t,...T\} $, the forward process of DDPM gradually adds Gaussian noise to each high-quality sub-volume of projection data $ \bm{y} $ extracted from $\bm{Y}$. The data distribution at each time step only depends on the previous state, satisfying the Markov property:
	
	\begin{equation}
		q(\bm{y}_t|\bm{y}_{t-1}) = \mathcal{N} (\bm{y}_t| \sqrt{1 - \beta_t}\bm{y}_{t-1}, \beta_t \bm{I}),
		\label{eq:1}
	\end{equation}
	where $ q(\bm{y}_{0}) = q(\bm{y}) $, and $ \bm{\beta} = \{\beta_1, \beta_2, \cdots, \beta_T \} $ is a predefined variance schedule. When the number of steps is large enough, the noisy sub-volume will gradually approach the standard normal distribution; i.e., $ \bm{y}_T  \sim \mathcal{N}(0,\bm{I}) $.
	By the nature of the Gaussian distribution, we can directly obtain the noised sub-volume at any timestamp:
	\begin{equation}
		\bm{y}_t =  \sqrt{\bar{\alpha}_t}\bm{y}_0 + \sqrt{1-\bar{\alpha}_t} \bm{\epsilon},
		\label{eq:2}
	\end{equation}
	where $\alpha_t = 1-\beta_t$, $ \bar{\alpha}_t = \prod_{i=1}^t \alpha_i $, and $ \bm{\epsilon}\sim \mathcal{N}(0, \bm{I}) $ is the noise to be added into the sub-volume.
	
	The reverse process starts from a pure noisy sub-volume $\bm{y}_T$ and moves backward step-by-step. At each step of the reverse process, the network is used to predict the noise $ \bm{\epsilon} $ added to the sum-volume $\bm{y}_t$, and then obtain the mean of the posterior distribution of $\bm{y}_{t-1}$, which is used as the denoised sub-volume. The process continues until it reaches the initial time step, producing the final denoised sub-volume $\bm{y}_0$. For projection data inpainting in this study, we can use the degraded sub-volume of projection data or image data into the network as the condition:
	
	\begin{equation}
		\bm{y}_{t-1} = 	\frac{1}{\sqrt{\bar{\alpha}_t}}\left(\bm{y}_t-\frac{1-\alpha_t}{\sqrt{1-\bar{\alpha}_t}}\bm{\epsilon}_{\bm{\theta}} (\bm{y}_t, \bm{z}, t)\right) +\sigma_t \bm{\xi},
		\label{eq:3}
	\end{equation}
	where $ \bm{\epsilon}_{\bm{\theta}} $ is the network parameterized with $ \bm{\theta} $ for predicting the noise $ \bm{\epsilon} $, $ \bm{z} $ represents a sub-volume of down-sampled projection data $\bm{Z}$, and $ \bm{\xi}\sim \mathcal{N}(0,\bm{I}) $ denotes the noise used for sampling.
	The loss function for training the noise prediction model is expressed as
	\begin{equation}
		\mathcal{L} = \mathbb{E}_{\bm{y}, \bm{z}}\mathbb{E}_{\bm{\epsilon},t} 	\left[ \left\| \bm{\epsilon} - \bm{\epsilon}_{\bm{\theta}} (\sqrt{\bar{\alpha}_t}\bm{y}_0 + \sqrt{1-\bar{\alpha}_t} \bm{\epsilon}, \bm{z}, t) \right\|_2^2 \right].
		\label{eq:11}
	\end{equation}
	
	The underlying principle of DDPM in the image domain is the same as that of DDPM in the projection domain. The high-quality sub-volume input $\bm{x}$ for the network is derived from the high-quality CT image $\bm{X}$, which is reconstructed from a full scan dataset. Meanwhile, the condition $\bm{u}$ is obtained from the degraded CT image $\bm{U}$, which is reconstructed from projection data inpainted by DDPM in the projection domain. Fig.~\ref{fig:1} illustrates the forward  and reverse diffusion processes of our conditional DDPM model that works synergistically in both the projection and image domains.
	
	\subsection{Parallel Pipeline}
\begin{figure*}[htbp]
		\centering
		\includegraphics[width=1.0\textwidth]{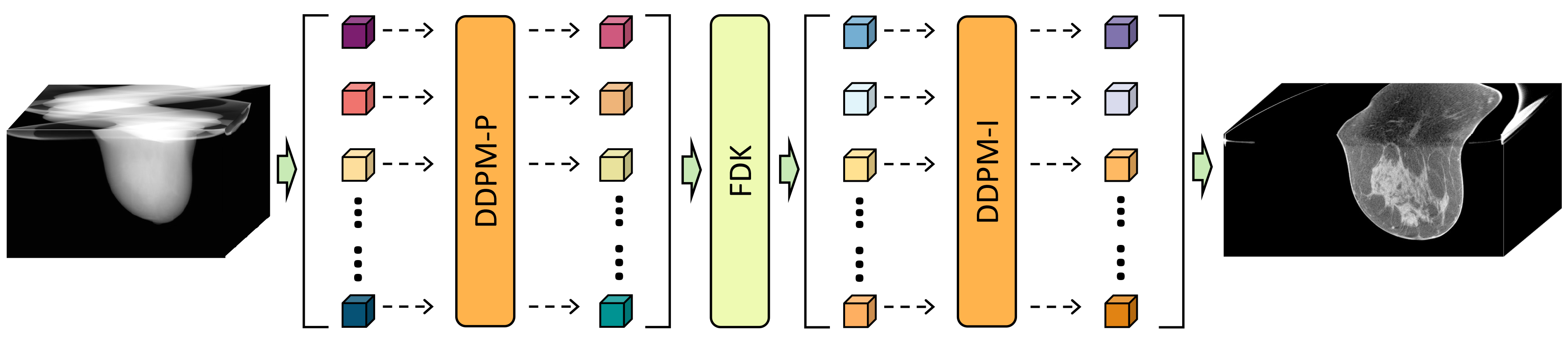}
		\caption{Conditional dual-domain DDPM model for sparse-view breast CT image reconstruction.}
		\label{fig:2}
	\end{figure*}

	As shown in Fig.~\ref{fig:2}, we trained two separate DDPMs for projection data and image data, denoted as DDPM-P and DDPM-I respectively. During the breast CT reconstruction from sparse-view data using the trained DDPMs, the projection data is initially divided into a number of 
	non-overlapped sub-volumes. Then, the missing data is restored using DDPM-P. In a distributed system, these sub-volume inpaintings are executed in parallel. Once all sub-volumes are repaired, each thread acquires the repaired results of all sub-volumes through data broadcasting. The sub-volumes are subsequently assembled into a full dataset. The repaired complete projection dataset is reconstructed using the FDK algorithm.
	
	Similar to the DDPM process in the projection data domain, the current image volume is divided into a number of non-overlapped sub-volumes, which are refined using DDPM-I in parallel. Finally, all sub-volumes are gathered and assembled into a complete high-quality image volume through data broadcasting. This entire workflow is intrinsically parallel, making sub-volume processing feasible on regular GPU and computationally more efficient. This pipeline can be implemented  on a cloud computing platform for deployment.
	
	\subsection{Dataset}
	
	The datasets for this study were acquired using a prototype cone-beam breast CT scanner (KBCT1000, Koning Corp., Norcross, GA, USA). This system is equipped with a tungsten-anode mammography-format X-ray tube (RAD71SP, Varex Imaging, Salt Lake City, UT, USA), operating at 49 kV with a 1.4 mm half-value layer (HVL). It also features an amorphous silicon flat-panel detector (PaxScan 4030CB, Varex Imaging, Salt Lake City, UT, USA). The detector panel array is composed of 768x1024 units, each measuring 388x388 \textmu m$^2$. The nominal image resolution is set at 1024x1024 pixels, with the number of slices varying according to the actual breast size, and each voxel size measuring 273.4$^3$ \textmu m$^3$.
	
	For this research, a total of 105 full-scan cone-beam datasets were collected. Of these, 84 were designated for the training set and the remaining 21 for testing purposes. The original full-scan dataset spanned an angular range of 360 degrees and consisted of 300 views. And we downsampled the data to 150 views for half-dose experiments and to 100 views for one-third dose experiments.
	
	\subsection{Training and Inference}
	
	The architecture of our Denoising Diffusion Probabilistic Model (DDPM) is an enhanced variant of the U-net model, as proposed by Ho et al. \cite{ho2020denoising}. This architecture is employed in both the DDPM for projection data (DDPM-P) and the DDPM for image data (DDPM-I), which are trained separately. Initially, DDPM-P is trained on full-scan projection data. We select random 16x16x16 sub-volumes from each dataset and pair these with corresponding sub-volumes from the sparse-view dataset to create pairs of fully-sampled and down-sampled data for training.
	
	The diffusion process in our model is set to 1,000 time steps, with a beta schedule linearly spaced between 1e-4 and 2e-2, following the guidelines suggested by Ho et al. \cite{ho2020denoising}. We use the AdamW optimizer with an initial learning rate of 1e-4, which is gradually reduced to 1e-5 over the course of training. The total number of training iterations for DDPM-P is set at 0.5 million.
	
	Upon completion of DDPM-P training, it is used to recover missing data in the training dataset. Subsequently, the Feldkamp-Davis-Kress (FDK) algorithm reconstructs an image volume from this data. However, the initial quality of these reconstructed images is typically suboptimal. To improve this, we train DDPM-I using pairs of these low-quality image volumes and their corresponding target high-quality volumes. The training methodology and parameters for DDPM-I mirror those used for DDPM-P.
	
	During the testing phase, the trained DDPMs are applied to the test dataset within a parallel computing pipeline. In both the projection and image domains, the size of the sub-volumes is uniformly set to 16x16x16, with a stride of (16,16,16) for extracting these sub-volumes.
	
	All training and inference processes were conducted on the Artificial Intelligence Multiprocessing Optimized System (AiMOS) cluster at the Center for Computational Innovations (CCI) at Rensselaer Polytechnic Institute, which is equipped with 1,576 Nvidia V100 GPUs.

	

	\section*{Acknowledgement}
	
	This research is partially supported by National Cancer Institute (NCI) of the National Institutes of Health (NIH) grants R21 CA134128 and R01 CA199044. The contents are solely the responsibility of the authors and do not necessarily reflect the official views of the NCI or NIH.
	
	\bibliography{ref}
\end{document}